\documentclass[conference]{IEEEtran}
\usepackage{amsmath}
\usepackage{amsfonts}
\usepackage{amssymb}
\usepackage{cite}
\usepackage{multicol}
\usepackage{verbatim}
\usepackage{graphicx}
\usepackage{balance}
\usepackage{epstopdf}
\usepackage{epsfig,color,algorithm}

\IEEEoverridecommandlockouts
\graphicspath{ {D:/images/} }

\begin{document}

\title{Enhancing Physical Layer Security in Dual-Hop  Multiuser  Transmission}
\author{\small Waqas Aman$^1$, Guftaar Ahmad Sardar Sidhu$^2$, Tayyaba Jabeen$^3$, Feifei Gao$^4$, and Shi Jin$^5$\\
$^1$ $^2$ $^3$ COMSATS Institute of Information Technology, Islamabad 44000, Pakistan\\
$^4$Tsinghua National
Laboratory for Information Science and Technology, Tsinghua
University, Beijing 100084, China\\
$^5$ National Communications Research Laboratory, Southeast University, Nanjing 210096, P. R. China\\
Emails: $^1$waqasaman87@gmail.com, $^2$guftaarahmad@comsats.edu.pk, $^3$tayyaba\_ jabeen92@yahoo.com}

\maketitle 

\begin{abstract}
In this paper, we consider the Physical Layer Security (PLS) problem in orthogonal frequency division multiple access (OFDMA) based dual-hop system which consists of multiple users, multiple amplify and forward relays, and an eavesdropper. The aim is to enhance PLS of the entire system by maximizing sum secrecy rate of secret users through optimal resource allocation under various practical constraints. Specifically, the sub-carrier allocation to different users, the relay assignments, and the power loading over different sub-carriers at transmitting nodes are optimized. The joint optimization problem is modeled as a mixed binary integer programming problem subject to exclusive sub-carrier allocation and separate power budget constraints at each node. A joint optimization solution is obtained through Lagrangian dual decomposition where KKT conditions are exploited to find the optimal power allocation at base station. Further, to reduce the complexity, a sub-optimal scheme is presented where the optimal power allocation is derived under fixed sub-carrier-relay assignment. Simulation results are also provided to validate the performance of proposed schemes.
\end{abstract}

\section{Introduction}
Broadcast nature of wireless communication provides many exciting opportunities, however makes the security of link a challenging issue.
Physical Layer Security (PLS) has gained much popularity to cope with physical link attacks. A wireless link is considered to be secure if it provides a positive non zero secrecy rate \cite{a}, and a link with  higher secrecy rate is known  as more secure link.  Orthogonal frequency division multiple access (OFDMA) has become a fundamental choice for next generation wireless communication networks because of its ability against multi path fading, high spectral efficiency, and provision of flexibility in resource allocation \cite{1}. To provide PLS in multi-carrier systems, resource optimization becomes the first choice and has been studied in \cite{17}-\cite{18}. In \cite{17}, two categories of users were considered: the secure users and the non secure users. The optimization seeks to maximize the throughput of non secure users via optimal power allocation subject to guaranteed average secrecy rate to secure users. The work in \cite{55} extended the previous work to maximizing secrecy rate in the presence of active eavesdropper which has capability to jam the secret user transmission. Further, the authors in \cite{18} considered multiple eavesdroppers and optimized sub-carrier assignment, power allocation, and secrecy data rate to  maximize the energy efficiency.

Dual-hop communication has recently gained significant attention in the field of wireless communication: it provides relay assisted communication which is used to enhance throughput, reduce power consumption, and to increase coverage area at the cell edges. To provide PLS in dual-hop networks, resource allocation has been widely studied under decode and forward (DF) relaying protocol  \cite{2}--\cite{5}. The authors in \cite{2} and \cite{3} studied the problem of optimal relay placement/assignment to enhance PLS. The work \cite{6} considered joint relay selection and power optimization to maximize the system's secrecy rate. Under OFDMA protocol, power allocation problem to maximize the secrecy rate was investigated in \cite{4}. The extension to this work with joint sub-carrier allocation and power loading was made in \cite{5}.  

The dual-hop transmission under amplify and forward (AF) protocols has become much attractive due to its simple implementation. 
However, the resource allocation in AF relay enhanced networks has always been a challenging task. Recently, different aspects of PLS in AF based systems has been studied in \cite{Huang}--\cite{Jendal}. The authors in \cite{Huang} investigated the impact of using an untrusted AF relay on secure communication and derived the exact Secrecy Outage Probability (SOP) under different transmission scenarios.  With mulitple trusted relays, \cite{LFan} proposed different relay selection strategies to enhance the PLS in multi-user cooperative relay networks.   The work \cite{Akhtar} focused on achievability of secrecy rate under different channel conditions. 
More recently,  the resource optimization in orthogonal frequency division multiplexing (OFDM) based single-user single-relay systems was considered in \cite{Jendal}. The authors studied the sub-carrier utilization and power allocation strategies under a total system power constraint and proposed a sub-optimal close form solution. The optimization under sum power constraint provides a good analysis of power allocation, however it may not be an attractive solution for practical systems. Moreover, under multi-user scenario a joint optimization of sub-carrier allocation and power allocation becomes necessary and is a challenging task.    

In this work, we maximize the sum secrecy rate of a multi-user multi-relay OFDM based dual hop network. We consider a joint optimization over a) sub-carrier allocation to users, b) relay assignment to secret users, and c) the power loading over sub-carriers under individual power constraints. A dual decomposition framework is adopted to derive the joint optimization solutions. 

\section{System Model}
We consider an OFDMA based multiple user multiple relay down-link transmission. We assume that all of the devices are equipped with single antenna and Channel State Information (CSI) of all the links is available at the base station (BS). We also assume that the eavesdropper is passive and targets the information of all the secret users. Further, we consider a dual-hop transmission mode where  direct links from the BS to the users and from the BS to the eavesdropper are missing due to large distance. Thus, all the secret users and the malicious node receive information messages only from the relay nodes which are operating under AF transmission protocol. Let $K$ and $J$ denote the total number of users and the total number of relays, respectively. An orthogonal transmission is assumed where total $N$ sub-carriers are available to  distribute among different users and relays under OFDMA protocol. 
\begin{figure}
\includegraphics[width=10cm, height=8cm]{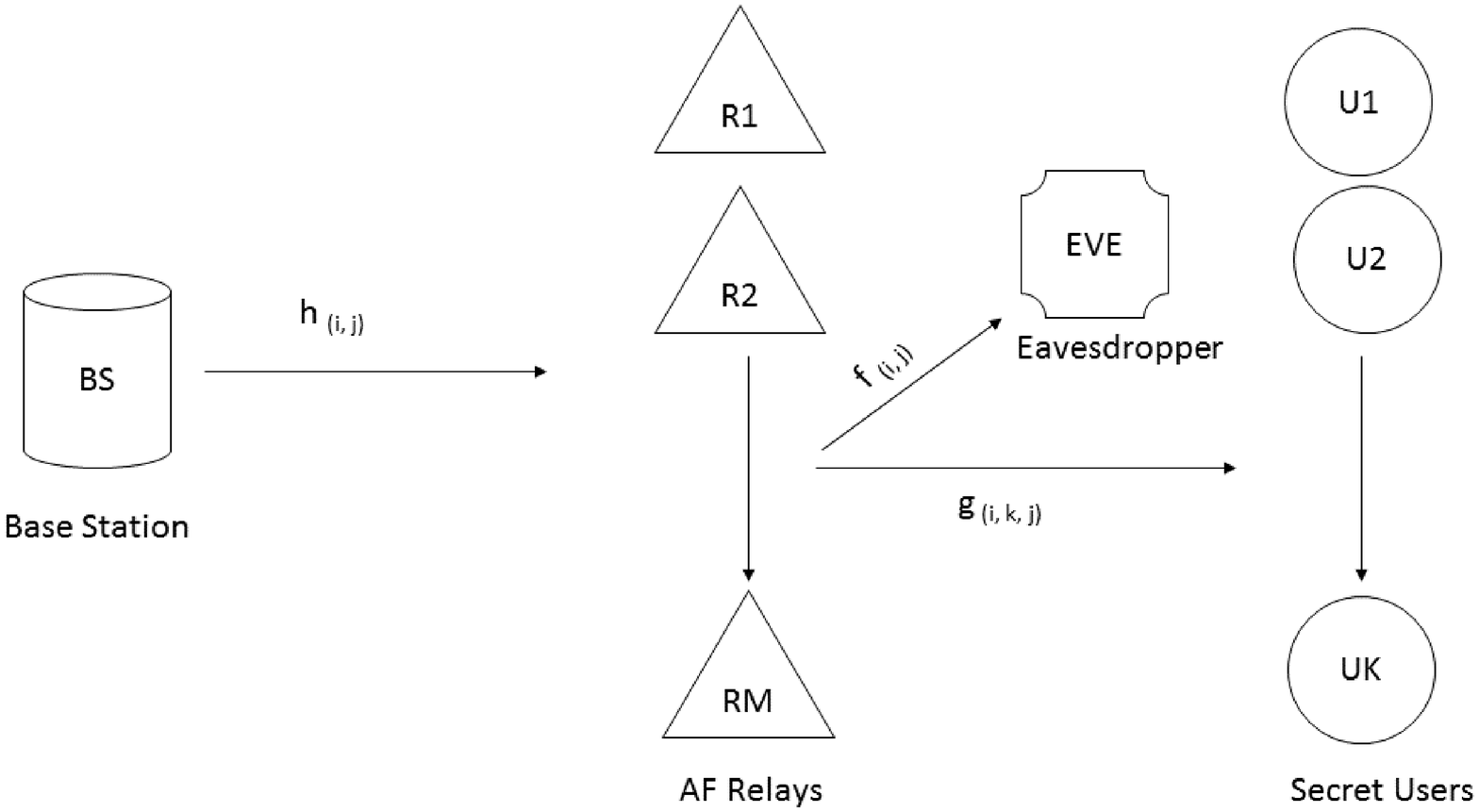}
\caption{System Model}
\label{Fig 1: System Model}
\end{figure}
 \\
The channel gain from BS to any $j$-th relay node and from the $j$-th relay to the $k$-th user on the $i$-th sub-carrier is denoted as $h_{i,j}$ and $g_{i,j,k}$ respectively. Similarly, the corresponding channel gain on the $i$-th sub-carrier from the $j$-th relay to the eavesdropper can be represented as $f_{i,j}$. Let, the $j$-th relay forward the received signal to the secret users in the presence of eavesdropper with amplification factor \textbf{$P_{i,j}$}:
\begin{align}
P_{i,j}=\sqrt{q_{i,j}/\left(p_i|h_{i,j}|^2+\sigma^2\right)},
\end{align}
where, $q_{i,j}$ is the $j$-th AF relay power on the $i$-th sub-carrier, $p_i$ is the BS transmit power on the $i$-th sub-carrier and $\sigma^2$ is the variance of Additive White Gaussian Noise (AWGN). Thus, the received Signal to Noise ratio (SNR) at the $k$-th secret user over the $i$-th sub-carrier from the $j$-th relay node is
\begin{align}
\text{SNR}_{i,j,k}^{SC}= \frac{P_{i,j}^2p_i|g_{i,j,k}|^2|h_{i,j}|^2}{{P_{i,j}^2|g_{i,j,k}|^2\sigma^2+\sigma^2}},
\end{align}
and the corresponding eavesdropper's SNR can be expressed as;
\begin{align}
\text{SNR}_{i,j}^{EV}= \frac{P_{i,j}^2p_i|f_{i,j}|^2|h_{i,j}|^2}{{P_{i,j}^2|f_{i,j}|^2\sigma^2+\sigma^2}}.
\end{align}

Assuming high SNR regime, we can express secrecy rate (SR) from the $j$-th relay to the $k$-th user on the $i$-th sub-carrier as;
\begin{align}
\text{SR}_{i,j,k}=&\frac{1}{2}\big[\log_2\left({1+\text{SNR}_{i,j,k}^{SC}}\right)-\log_2\left({1+\text{SNR}_{i,j}^{EV}}\right)\big],\nonumber \\ 
%
=&\frac{1}{2}\log_2\left(\frac{b_{i,j,k}\left({1+a_{i,j}p_i+q_{i,j}c_{i,j}}\right)}{\left({1+a_{i,j}p_i+q_{i,j}b_{i,j,k}}\right)c_{i,j}}\right),
\end{align}
where $a_{i,j}=\frac{|h_{i,j}|^2}{\sigma^2}$, $b_{i,j}=\frac{|g_{i,j,k}|^2}{\sigma^2}$, and $c_{i,j}=\frac{|f_{i,j}|^2}{\sigma^2}$. The $\frac{1}{2}$ factor appears in the above expression as a complete transmission from BS to users takes two time slots.

We adopt a fully flexible relay and sub-carrier allocation strategy where a relay can be allocated to more than one users, and each user can be served with multiple relay nodes over different sub-carriers. Further a sub-carrier is allocated to the same user over two hops of transmission. On account of sub-carrier allocation and relay selection, we define two binary  variables:  $\alpha_{i,k} \in {[0,1]}$ such that $\alpha_{i,k}=1$ when the $i$-th sub-carrier is allocated to the $k$-th user and zero otherwise, and $\beta_{i,j} \in {[0,1]}$ such that $\beta_{j,k}=1$ when the $j$-th relay is allocated to the $k$-th user. With this the sum SR of the system can be expressed as;
\begin{align}
\text{SR}_{\text{sum}}=\sum_{i=1}^N\sum_{j=1}^J\sum_{k=1}^K \alpha_{i,k}\beta_{j,k} \text{SR}_{i,j,k}.
\end{align}

\section{Joint Optimization Problem and Proposed Schemes}
The focus of this work is to maximize SR$_{\text{sum}}$ with jointly optimizing the relay selection, sub-carrier assignment and BS's transmit power loading over different sub-carriers. For simplicity, we adopt a uniform power allocation strategy at the relay nodes such that total power allocated at each relay does not exceeds the maximum power limits, i.e., 
$\sum_{i=1}^N q_{i,j}\leq Q_j, \ \ \forall j,$ 
where $Q_{j}$ is total available power budget at the $j$-th relay. Let $N_j$ denote the number of sub-carriers allocated to the $j$-th relay, then the power allocated at the $i$-th sub-carrier is $q_{i,j}=\frac{Q_j}{N_j}$. Let $PT$ be the total power available at BS. Then the joint sub-carrier allocation, relay assignment, and power loading optimization problem can be formulated as:
\begin{align}
\max_{(p_{i},\alpha_{i,k},\beta_{j,k})}\quad &\ \text{SR}_{\text{sum}} \\
\text{s.t.} \ \ \quad &\sum_{k=1}^K\alpha_{i,k}= 1, \nonumber \quad \forall i,
\\&\sum_{j=1}^J\beta_{j,k}\leq J, \nonumber  \; \;  \forall k, \quad \sum_{k=1}^K\beta_{j,k}\leq K, \; \; \forall j,
\\&\sum_{i=1}^N\sum_{j=1}^J\sum_{k=1}^K \alpha_{i,k}\beta_{j,k}p_i \leq PT. \nonumber
\end{align}
\\
The first constraint ensures that one sub-carrier can not be assigned to more than one user while the next two constraints depict that a relay can be allocated to more than one users and a particular user may exploit multiple relays for secure transmission. The last constraint represents that sum transmit power on all sub-carriers at BS should be less than or equal to a maximum power limit.

\subsection{Joint Optimization Scheme}\label{sec:jntopt}
The problem (6) is mixed binary integer programming problem, and a vast search over all variables is needed to find an optimal solution. Thanks to \cite{8}, the difference between the solution of dual problem and the solution of primal problem\footnote{Commonly known as duality gap.}   becomes zero when we have sufficiently large number of sub-carriers in OFDM based transmission regardless of convexity of original problem. The dual problem associated with primal problem (6) can be defined as:
\begin{align}
\min_{\lambda\geq0} \quad D(\lambda)
\end{align}
where $\lambda$ is a dual variable, and the dual function $D(\lambda)$ can be expressed as
\begin{align}
D(\lambda)=\max_{(p_{i},\alpha_{i,k},\beta_{j,k})} \quad &L(p_{i},\alpha_{i,k},\beta_{j,k})\\
\text{s.t.}\ \ \quad &\sum_{k=1}^K \alpha_{i,k} = 1 \nonumber ,\quad \forall i,\\
&\sum_{j=1}^J\beta_{j,k}\leq J, \nonumber  \  \forall k, \quad \sum_{k=1}^K\beta_{j,k}\leq K, \ \forall j,
\end{align}
with,
\begin{align}
L(p_{i},\alpha_{i,k},\beta_{j,k})=&\sum_{i=1}^N\sum_{j=1}^J\sum_{k=1}^K \alpha_{i,k}\beta_{j,k} \text{SR}_{i,j,k}+\\
&\lambda\left(PT-\sum_{i=1}^N\sum_{j=1}^J\sum_{k=1}^K \alpha_{i,k}\beta_{j,k}p_i\right). \nonumber
\end{align}

To solve the dual problem we first solve the dual function $D(\lambda)$ and  similar to \cite{9} we solve the problem through dual decomposition approach. The problem can be rewritten as
\begin{align}
D(\lambda)=\max_{(p_{i},\alpha_{i,k},\beta_{j,k})} \quad &\sum_{i=1}^N\sum_{j=1}^J\sum_{k=1}^K \alpha_{i,k}\beta_{j,k}(SR_{i,j,k}-\lambda p_i) \\ 
\text{s.t.} \quad &\sum_{k=1}^K\alpha_{i,k}= 1 \nonumber, \quad \forall i,\\
&\sum_{j=1}^J\beta_{j,k}\leq J, \nonumber  \  \forall k, \quad \sum_{k=1}^K\beta_{j,k}\leq K, \ \forall j.
\end{align}
For any given sub-carrier allocation and relay assignment, the optimal power allocation can be obtained from
\begin{align}
\max_{p_i \geq 0} \ \ \log_2\left(\frac{b_{i,j,k}(1+a_{i,j}p_i+q_{i,j}c_{i,j})}{(1+a_{i,j}p_i+q_{i,j}b_{i,j,k})c_{i,j}}\right)-\lambda p_i.
\end{align}
The problem is a convex optimization and close form solution can be obtained exploiting the standard techniques. Let $p_i^*$ denote the obtained optimal power allocation over the $i$-th sub-carrier. The dual function becomes
\begin{align}
D(\lambda)=\max_{(\alpha_{i,k},\beta_{j,k})} \quad &\sum_{i=1}^N\sum_{j=1}^J\sum_{k=1}^K \alpha_{i,k}\beta_{j,k}(\text{SR}_{i,j,k}^*-\lambda p_i^*) \\ 
\text{s.t.} \quad &\sum_{k=1}^K\alpha_{i,k}= 1 \nonumber, \quad \forall i,\\
&\sum_{j=1}^J\beta_{j,k}\leq J, \nonumber  \  \forall k, \quad \sum_{k=1}^K\beta_{j,k}\leq K, \ \forall j,
\end{align}
where $\text{SR}_{i,j,k}^*$ is obtained by putting value of $p_i^*$ into (11). Now, we need to find the optimal sub-carrier allocation and relay selection. For immediate recovery of the binary variables $\alpha_{i,k}$ and $\beta{j,k}$, we define a new variable $\eta_{i,j,k} \in \{0,1\}$ such that $\eta_{i,j,k}=1$ if $\alpha_{i,k}\beta_{j,k}=1$ and zero otherwise. The problem (12) can be rewritten as 
\begin{align}
D(\lambda)=\max_{\eta_{i,j,k}} \quad &\sum_{i=1}^N\sum_{j=1}^J\sum_{k=1}^K \eta_{i,j,k}(\text{SR}_{i,j,k}^*-\lambda p_i^*)\\
\text{s.t.}\quad &\sum_{k=1}^K\sum_{j=1}^J\eta_{i,j,k}= 1 \nonumber,\quad \forall i.
\end{align}
The optimum solution of above problem is to assign a sub-carrier relay pair $(i,j)$ to user $k$ which maximizes the $\text{SR}_{i,j,k}^*$, i.e.,  
\begin{align}
(i^*,j^*,k^*)=\arg \max_{i} \ \ \text{SR}_{i,j,k}^*,
\end{align}

\begin{align}
\eta_{i,j,k}^*&=\begin{cases}
1, &  (i^*,j^*,k^*)=(i,j,k) \nonumber \\
\\
0, &  \text{otherwise}.
\end{cases}
\end{align}
Now the optimum sub-carrier allocation and relay assignment are obtained. Let $\alpha_{i,k}^*$ and $\beta_{j,k}^*$ denote the optimal assignment variables. Thus, substituting $p_i*$, $\alpha_{i,k}^*$, and $\beta_{j,k}^*$ in (9) we obtain the dual function.  

Next we solve the dual problem (7) and use the sub gradient method to find the optimum value of the dual variable $\lambda$. The $\lambda$ is updated iteratively according to the following  updates;
\begin{align}
\lambda(t+1)= \lambda(t)+\delta(t) \left(\sum_{i=1}^N\sum_{j=1}^J\sum_{k=1}^K \alpha_{i,k}^*\beta_{j,k}^*p_i-PT\right),
\end{align}
where $t$ represents the $t$-th iteration of iterative update and $\delta$ is the step size. In each update of $\lambda$, the optimum power allocation, the sub-carrier assignment and relay selection variables are updated. At convergence, the optimum values of both dual and the primal variables are obtained.

It remains to find the solution of problem in (11). The Lagrangian associated with the optimization is
\begin{align}
\Delta=
\log_2 \left({\frac{b_{i,j,k}({1+a_{i,j}p_i+q_{i,j}c_{i,j}})}{{(1+a_{i,j}p_i+q_{i,j}b_{i,j,k}})c_{i,j}}}\right)- \lambda p_i+\sum_{i=1}^N \gamma_i p_i,
\end{align}
where $\gamma_i$ is the Lagrangian multiplier. By setting $ \frac{\partial \Delta}{\partial p_i}=0 $ 
we have
\begin{align}
\gamma_i=  \lambda-\frac{a_{i,j}b_{i,j,k}^2c_{i,j}q_{i,j}-a_{i,j}b_{i,j,k}c_{i,j}^2q_{i,j}}{b_{i,j,k}(\zeta_{i,j}+q_{i,j}c_{i,j})(\zeta_{i,j}+q_{i,j}b_{i,j,k})c_{i,j}},
\end{align}
where $\zeta_{i,j}=1+a_{i,j}p_i$.
Now applying KKT condition $\gamma_i p_i=0$, we obtain
\begin{align}
p_i \left(\lambda-\frac{a_{i,j}b_{i,j,k}^2c_{i,j}q_{i,j}-a_{i,j}b_{i,j,k}c_{i,j}^2q_{i,j}}{b_{i,j,k}(\zeta_{i,j}+q_{i,j}c_{i,j})(\zeta_{i,j}+q_{i,j}b_{i,j,k})c_{i,j}}\right)=0.
\end{align}
For $p_i>0$, there is 
\begin{align}
\lambda=\frac{a_{i,j}b_{i,j,k}^2c_{i,j}q_{i,j}-a_{i,j}b_{i,j,k}c_{i,j}^2q_{i,j}}{b_{i,j,k}(\zeta_{i,j}+q_{i,j}c_{i,j})(\zeta_{i,j}+q_{i,j}b_{i,j,k})c_{i,j}}.
\end{align}
Solving for $p_i$, it results in $p_i^*$:   
\begin{align}
p_i^*= \left(\frac{-B\pm\sqrt{B^2-4AC}}{2A}\right)^+,
\end{align}
where \\ $A=a_{i,j}^2b_{i,j,k}c_{i,j}$.\\ \quad $B=2a_{i,j}b_{i,j,k}c_{i,j}+a_{i,j}b_{i,j,k}^2c_{i,j}q_{i,j}+a_{i,j}b_{i,j,k}^2c_{i,j}q_{i,j}$.\\ \quad 
$C=b_{i,j,k}c_{i,j}+b_{i,j,k}^2c_{i,j}q_{i,j}+b_{i,j,k}c_{i,j}^2q_{i,j}+b_{i,j,k}^2c_{i,j}^2q_{i,j}^2- \frac{a_{i,j}b_{i,j,k}^2c_{i,j}q_{i,j}}{\lambda}+\frac{a_{i,j}b_{i,j,k}c_{i,j}^2q_{i,j}}{\lambda}$. \\ \\
This completes the joint optimization solution.

\subsection{Sub Optimal Scheme}\label{sec:subopt}
The proposed solution in section \ref{sec:jntopt} provides a joint optimization over different variables. In this section, to reduce the computational complexity, we propose a sub-optimal scheme where only the power allocation is optimized for a fixed sub-carrier allocation. The steps involved in the algorithm are listed as follows:
\begin{itemize}
\item[1.] Randomly allocate all the sub-carriers such that the $i$-th sub-carrier is exclusively allocated to a unique user-relay pair $(j,k)$. Thus, both $\alpha_{i,k}$, and $\beta_{j,k}$ are obtained.
\item[2.] Based on the sub-carrier allocation to each node, find the relay power allocation under uniform power distribution scheme.   
\item[3.]With obtained sub-carrier and relay allocation, the power allocation problem reduces to single user single relay power allocation problem, i.e.,  
\begin{align}
\max_{p_{i}}\quad &\ \sum_{i=1}^N\sum_{j=1}^J\sum_{k=1}^K \alpha_{i,k}\beta_{j,k} \text{SR}_{i,j,k} \\
\text{s.t.} \quad  &\sum_{i=1}^N\sum_{j=1}^J\sum_{k=1}^K \alpha_{i,k}\beta_{j,k}p_i \leq PT. \nonumber
\end{align}
This problem can be solved using dual decomposition with similar steps described in previous subsection. However note that now we need to find only $N$ power variables instead of $NKJ$ variables for power allocation at BS. The exact solution is missing for simplicity.  
\end{itemize}  
 
\section{Simulation Results}
This section includes simulation results to validate the performance of proposed algorithms. Following three schemes are compare:
\begin{itemize}
\item OPT: It represents the joint optimization solution where the power allocation over different sub-carriers at BS, sub-carrier assignment to different users, and relay selection for each of the user are found simultaneously, as 
given in section \ref{sec:jntopt}. 
\item Sub-OPT: This scheme finds the optimum power distribution for the fixed sub-carrier and relay allocation as given in section \ref{sec:subopt}. 
\item Non-OPT: This denotes a trivial solution without optimization where the sub-carriers are randomly allocated to different users and relays and then available power is distributed evenly over all the allocated sub-carriers.
\end{itemize}
\begin{figure}
\includegraphics[width=9.5cm, height=9cm]{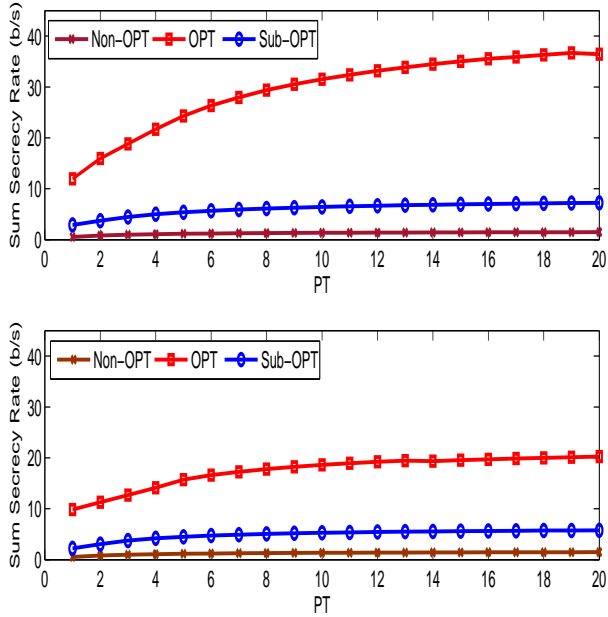}
\caption{Sum Secrecy Rate vs Total Transmit Power (PT)}
\label{Fig 2: Sum Secrecy Rate vs Total Transmit Power (PT)}
\end{figure}
We choose 6 tap channels taken from  i.i.d Gaussian random  variables for all links and assume the same noise variance at all nodes. The figure of merit is taken as the sum secrecy rate of all secret users. A total of $K=12$ secrete users, a single eavesdropper,  and $J=4$ relays are considered for the simulation.  

Figure 2 shows the sum secrecy rate versus the total transmit power $PT$ where the transmit power is in absolute form and varies from $1$ to $20$. To look into the effect of the number of sub-carriers $N$, we plotted the results for $N=64$ (upper subplot) and $N=32$ (lower subplot). It can be observed that the sum secrecy rate increases by increasing the total power $PT$. The OPT significantly outperforms as compared to the other two candidates. The Sub-OPT also gives considerable gain over the Non-OPT solution. Further, at higher values of $PT$ the gap between the OPT and the other two schemes increases. Moreover, the gap between the OPT and other two schemes also increases with the increase in the number of sub-carriers from 32 to 64. On the other hand, the Sub-OPT exhibits similar performance at different $N$. The performance gap between Sub-OPT and Non-OPT becomes almost constant from $PT=8$ onward.  

\begin{figure}
\includegraphics[width=9.5cm, height=9cm]{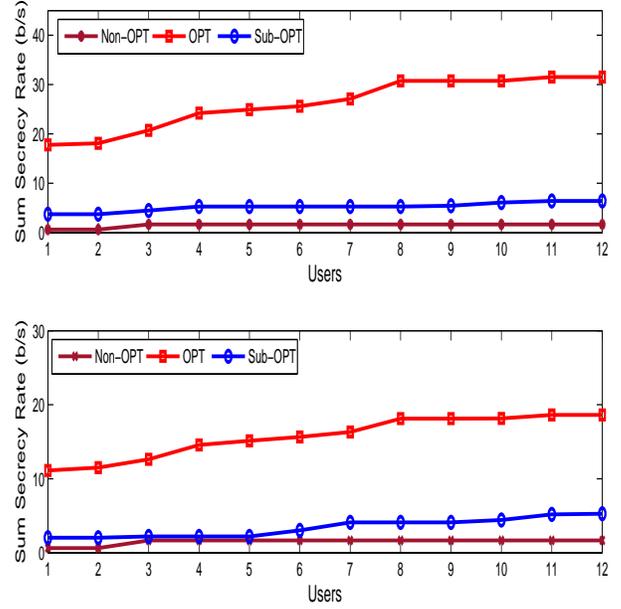}
\caption{Sum Secrecy Rate vs Total Number of Secret Users(K)}
\label{Fig 3: Sum Secrecy Rate vs Total Transmit Power (PT)}
\end{figure}
Next, we fix the total transmit power at $PT=10$ for the same scenario explained above and evaluate the results by varying number of secret users, i.e., $K= \ 1 \ \text{to} \ 12$.  Similar trends in the performance are observed. Like the previous figure, the upper and the lower subplots carry results for $N=64$ and $N=32$, respectively.  With the higher number of users, higher sum secrecy rate is achieved. This is because multi-user diversity effect increases with the increase in the number of users, that is to say each sub-carrier has more chances to get a better user-relay pair. Further, increasing the number of sub-carriers also results in higher secrecy rate. It is interesting to note that when the total number of users are less, the Sub-OPT does not provide significant gain over Non-OPT. However, either increasing the number of users or the number of sub-carrier, the Sub-OPT provides prominent advantage over the Non-OPT. On the other side, the proposed joint optimization scheme `OPT' always provide a considerable performance gain over the trivial solutions.   
\section{Conclusions} \label{sec:con}
This work considered resource optimization for PLS in multi-user based dual hop relay network. A joint optimization problem was formulated to maximize the sum secrecy rate through the optimal sub-carrier allocation, relay assignment, and power loading at BS. Convex optimization techniques were used to find an efficient joint optimization solution under OFDMA based sub-carrier allocation constraints and power constraints. Further, a sub-optimal solution was also presented which provides considerable gain over the non optimized solution. Simulation results showed that the proposed scheme outperforms the trivial solutions and exhibits significant performance enhancement at different values of $PT$ and $K$. 
It is observed from the results that the gain of joint optimization solution increases with the increase in the number of sub-carriers.


\begin{thebibliography}{100}
\bibitem{a}
A. D. Wyner, ``The wire-tap channel," in \textit{  The Bell Sys. Tech. Journal}, vol. 54, no. 8, pp. 1355--1387, Oct 1975.
\bibitem{1}
D. W. K. Ng and R. Schober, ``Resource allocation for secure OFDMA communication systems," in \textit{proc. IEEE Australian Commun. Theory Workshop (AusCTW)}, Melbourne, pp. 13--18, May 2011.
\bibitem{17}
X. Wang, M. Tao and Y. Xu, ``Power and subcarrier allocation for physical-layer security in OFDMA based broadband wireless networks," \textit{IEEE Trans. on Info. Forensics and Security}, vol. 6, no. 3, pp. 693--702, Sept 2011.
\bibitem{55}
M. R. Javan and N. Mokari,``Resource allocation for maximizing secrecy rate in presence of active eavesdropper," in \textit{proc. IEEE 22nd Iranian Conf. on Electrical Engineering (ICEE 2014)}, pp. 1565--1568, Tehran, 2014.
\bibitem{18}
D. W. K. Ng, E. S. Lo and R. Schober, ``Energy-efficient resource allocation for secure OFDMA systems," \textit{IEEE Trans. on Vehicular Tech.}, vol. 61, no. 6, pp. 2572--2585, May 2012.
\bibitem{2}
J. Mo, M. Tao and Y. Liu, ``Relay placement for physical layer security: A secure connection perspective," in \textit{proc. IEEE Commun. Letters}, vol. 16, no. 6, pp. 878--881, April 2012.
\bibitem{3}
H. Deng, H. Ming Wang, W. Wang and Q. Yin, ``Secrecy transmission with a helper: To relay or not to relay," in \textit{proc. IEEE Int. Conf. on Commun. Workshops (ICC)}, Sydney, pp. 825--830, 2014.

\bibitem{6}
C. Wang, H.-M. Wang and X.-G. Xia, ``Hybrid opportunistic relaying and jamming with power allocation for secure cooperative networks," \textit{IEEE Trans. on wireless commun.}, vol. 14, no. 2, pp. 589--605, September 2014.
\bibitem{4}
C. Jeong, and I.-M. Kim,``Optimal power allocation for secure multicarrier relay system." \textit{IEEE Trans. on Signal Processing}, vol. 59, no. 11, pp. 5428--5442, July 2011.
\bibitem{5}
A. Wang, J. Cheng et al, ``Joint subcarrier and power allocation for physical layer security in cooperative OFDMA network," \textit{EURASIP J. on Wireless Commun. and Networking}, pp. 1--10, 2013.

\bibitem {Huang}
J. Huang, A. Mukherjee and A. lee, ``Secure communication via an untrusted non-regenerative relay in fading channels," \textit{IEEE Trans. on Signal Processing}, vol. 61, no. 10, pp. 2536-2551, 2013.  
\bibitem {LFan}
L. Fan, X. Lie, Q. Trung, and M. Elkashlam, ``Secure multiuser communications in multiple amplify-and-forward relay networks",  \textit{IEEE Trans. on commun.}, vol. 62, no. 9, pp. 3299--3310, September 2014.
\bibitem {Akhtar}
A. M. Akhtar, A. Behnad and X. Wang, ``On the secrecy rate achievability in dual-hop amplify-and-forward relay networks," \textit{IEEE Wireless Commun. Letters}, vol. 3, no. 5, pp. 493-496, 2014. 
\bibitem {Jendal}
A. Jendal and R. Bose, ``Resource allocation for secure multicarrier AF relay system under total power constraint," \textit{IEEE Commun. Letters}, vol. 19, no. 2, pp. 231-234, Feb 2015.

\bibitem{8}
W. Yu and R. Lui, ``Dual methods for nonconvex spectrum optimization of multicarrier systems," \textit{IEEE Trans. on Commun.}, vol. 54, no. 7, pp. 1310--1322, July 2006.
\bibitem{9}
G. A. S. Sidhu, F. Gao, W. Wang, and W. Chen, ``Resource allocation in relay-aided OFDM cognitive radio networks," \textit{IEEE Trans. on Vehicular Tech.}, vol. 62, no. 8, pp. 3700-3710, Oct. 2013.
\end{thebibliography}
\end{document}